\documentclass[aps,prb,twocolumn,showpacs,floats]{revtex4}
\usepackage[dvips]{graphicx}
\usepackage{amsmath}
\usepackage{epsfig}
\usepackage{psfrag}
\newcommand{\be}{\begin{equation}}
\newcommand{\ee}{\end{equation}}
\newcommand{\bea}{\begin{eqnarray}}
\newcommand{\eea}{\end{eqnarray}}
\newcommand{\ba}{\begin{array}}
\newcommand{\ea}{\end{array}}
\newcommand{\nn}{\nonumber}

\newcommand{\del}{\delta}

\newcommand{\gam}{\gamma}
\newcommand{\al}{\alpha}
\newcommand{\sig}{\sigma}

\newcommand{\noi}{\noindent}
\newcommand{\eps}{\epsilon}
\newcommand{\lam}{\lambda}

\newcommand{\ria}{\rightarrow}
\begin{document}

\title{\bf Quantum pumping in a ballistic graphene bilayer}

\author{
G.M.M. Wakker and M. Blaauboer
}

\affiliation{Kavli Institute of Nanoscience, Delft University of Technology,
Lorentzweg 1, 2628 CJ Delft, The Netherlands
}
\date{\today}

\begin{abstract}
We investigate quantum pumping of massless Dirac fermions in an ideal (impurity free) 
double layer of
graphene. The pumped current is generated by adiabatic variation 
of two gate voltages in the contact regions to a weakly doped double graphene sheet.
At the Dirac point and for a wide bilayer with width $W$ $\gg$ length $L$, we find
that the pumped current scales linearly with the interlayer coupling length $l_{\perp}$ 
for $L/l_{\perp} \ll 1$, is maximal for $L/l_{\perp} \sim 1$, and crosses over to a 
$\ln(L/l_{\perp})/(L/l_{\perp})$-dependence for $L/l_{\perp} \gg 1$. We compare
our results with the behavior of the conductance in the same system and discuss their 
experimental feasibility.
\end{abstract}

\maketitle

\section{Introduction}

Quantum pumping of charge refers to the generation of a dc electrical current in 
the absence of an applied bias voltage by periodic (ac) modulation of 
two or more system parameters, for example the shape of the confining potential or 
a magnetic field~\cite{alts99}. 
The idea of adiabatically generating a flow of particles
in a moving periodic potential is due to Thouless~\cite{thou83}, and has been followed by many
theoretical and a few experimental investigations of pumping in mesoscopic systems. 
In 1998 Brouwer~\cite{brou98}, building on earlier results by B\"{u}ttiker 
{\it et al.}~\cite{buet94} and a proposal
by Spivak {\it et al.}~\cite{spiv95}, developed a 
description of quantum pumping through open mesoscopic systems in terms of
the scattering matrix of the system. This paved the way for investigating the effects of
quantum interference on quantum pumping and has led to theoretical investigations of
many different aspects of pumped currents in open nanodevices, such as 
the relation of quantum pumping to geometric 
(Berry) phases~\cite{avro00}, the effect of Andreev reflection on quantum pumping 
in nanostructures that contain superconducting parts~\cite{wang01}, the 
effect of electron-electron interactions~\cite{sple95}, and the generation of 
adiabatically pumped spin currents~\cite{mucc02}.

Most of these investigations were carried out for semiconductor structures such as quantum dots
and nanowires. 
In addition, also proposals for adiabatic pumping in carbon nanotubes~\cite{wei01}, 
and recently graphene~\cite{prad09,zhu09} have been put forward. Since its
experimental discovery in 2004~\cite{novo04}, graphene has been found to exhibit electronic 
transport properties that are quite different from those in other nanoelectronic structures due
to the nature of its charge carriers (massless Dirac fermions described by a relativistic wave 
equation)~\cite{cast09}. An example of this is the importance of 
evanescent modes for transport close to the Dirac point: in a sample of undoped graphene,
which does not have any free electrons, contacted by doped electrodes the conductance close to the Dirac point 
is dominated by the contribution of evanescent modes which transmit electrons injected from one end of 
the sample to the other end~\cite{kats06,twor06}. This is also true for quantum pumping in a monolayer
of graphene, where the pumped 
current is induced by two 
oscillating gate voltages~\cite{prad09}. 

In this paper we investigate quantum pumping of Dirac fermions in a graphene bilayer. Compared to a carbon
monolayer, the bilayer has an additional energy scale, namely the interlayer coupling strength $t_{\perp}$. 
The corresponding lengthscale $l_{\perp}$ is an order of magnitude larger than the interatomic distance 
$d$~\cite{cast09}.
Our aim
is to investigate the dependence of the adiabatically pumped current on the interlayer coupling $t_{\perp}$.
For the conductance in a bilayer with heavily doped contact regions and width $W$ $\gg$ length $L$ it has 
recently been found that at the Dirac point the bilayer transmits as 
two monolayers in parallel and the conductance is independent of $t_{\perp}$~\cite{snym07}.
In contrast, for the pumped current $I_p$ we find that $I_p$ depends linearly on $L/l_{\perp}$ for
small interlayer coupling strength $t_{\perp}=1/l_{\perp} \ll 1/L$, exhibits a maximum around
$t_{\perp} \sim 1/L$ and scales as $\ln(L/l_{\perp})/(L/l_{\perp})$ for large interlayer coupling
$t_{\perp}=1/l_{\perp} \gg 1/L$. For typical experimental parameters the pumped current is
of order 10-100 pA.

The paper is organized as follows. In Sec.~\ref{model1} we present the bilayer model and an introduction to 
pumped currents. Sec.~\ref{calculations} contains the calculation of the pumped current, 
followed by results and comparison to the conductance in Sec.~\ref{results}.  
We conclude by making a connection to experiments in Sec.~\ref{conclusion}.

\section{Bilayer model and pumped current} 
\label{model1}

\begin{figure}[t]
\centering
\vspace{-0.6in}
\includegraphics[width=\columnwidth]{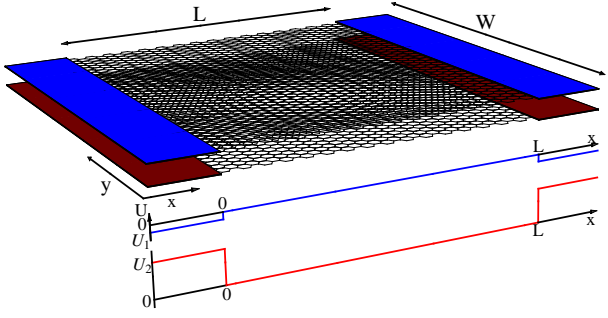}
\vspace{-0.6in}
\caption[]{Schematic picture of the graphene bilayer. Top panel: Two stacked honeycomb lattices of carbon atoms in a strip of width $W$
between metallic contacts (blue and red regions). Bottom panel: Variation of the electrostatic potential across the two layers.
} 
\label{fig:geometry}
\end{figure}

We consider the geometry that is schematically depicted in Fig.~\ref{fig:geometry}. A sheet of ballistic graphene
in the $(x,y)$-plane contains a weakly doped strip of length $L$ and width $W$ which is contacted by two 
more heavily doped electrodes at $x=0$ and $x=L$. The doping in these contacts is controlled by gate voltages, 
which induce a potential 
profile of the form
\be
U(x) = \left\{ \ba{ll}
U_1(t) & \ \mbox{\rm for} \ x<0\ \mbox{\rm or}\ x>L\ \ \mbox{\rm in the upper layer} \\
U_2(t) & \ \mbox{\rm for} \ x<0\ \mbox{\rm or}\ x>L\ \ \mbox{\rm in the lower layer} \\
0 & \ \mbox{\rm for} \ 0<x<L \ \ \mbox{\rm in both layers}
\ea \right.
\ee 

We assume the potential step to be abrupt, which is justified close to the Dirac point where the 
Fermi wavelength $\lam_F$ $\gtrsim$ $L$ and any smoothing of the step over a distance small compared
to $L$ becomes negligible. In addition, we consider a short and wide geometry ($L \ll W$), for which 
boundary conditions in the transverse $y$-direction become irrelevant. 

The bilayer pump is operated by periodic variations of the carrier density in the leads by varying the potentials 
$U_1(t) = U_1 + \delta U_1 \cos (\omega t)$
and $U_2(t) = U_2 + \delta U_2 \cos (\omega t + \phi)$ such that $U_1(t)$ = $U_2(t)$ on average. In the linear response 
regime where $\delta U_i \ll U_i$ ($i=1,2$), the pumped current $I_p$ into the left lead is given by the scattering 
matrix expression~\cite{brou98} 
\begin{equation}
I_p \equiv I_L = \frac{\omega e \sin \phi\, \delta U_1 \delta U_2}{2 \pi} \sum_{\alpha \in L} \sum_{\beta}\ \mbox{\rm Im}\ 
\left( \frac{\partial S_{\alpha \beta}^{ \star}}{\partial U_1} \frac{\partial S_{\alpha \beta}}{\partial U_2} \right).
\label{eq:current}
\end{equation}  
Here the index $\al$ sums over all modes in the left contact region and $\beta$ sums over all modes in both the 
left and right contact regions. The pumped current into the right lead is then given by $I_R = - I_L = -I_p$.
$S$ denotes the Landauer-B\"uttiker scattering matrix whose elements $S_{\al \beta, nm}$ describe scattering from mode $m$ in 
lead $\beta$ to mode $n$ in lead $\al$.

In the presence of a potential $U_1$ ($U_2$) in the upper (lower) layer, the low-energy excitations 
of the graphene bilayer close to a Dirac point are descibed by the 4$\times$4 Hamiltonian, $H=$ 
\begin{eqnarray}
\label{Ham}
H \hspace{-0.03in} = \hspace{-0.03in} \left( \begin{array}{cccc}
U_1 & \hspace{-0.05in} v(p_x + i p_y) \hspace{-0.05in} & t_{\perp} & 0 \\
\hspace{-0.05in} v(p_x - i p_y) \hspace{-0.05in} & U_1 & 0 & 0 \\
t_{\perp} & 0 & U_2 & \hspace{-0.05in} v(p_x - i p_y) \hspace{-0.05in} \\
0 & 0 & \hspace{-0.05in} v(p_x + i p_y) \hspace{-0.05in} & U_2 \end{array} \right), 
\end{eqnarray} 
where $\bf{p}$ $=$ $-i \hbar \partial / \partial \bf{r}$ is the momentum operator. The Hamiltonian (\ref{Ham}) acts
on the four-component wavefunction ($\psi_{A_1}$, $\psi_{B_1}$, $\psi_{B_2}$, $\psi_{A_2}$), with $A_1$ labeling 
the amplitude on the $A$-sublattice of the first (upper) layer, and similarly for $B_1$, $A_2$ and $B_2$. 
We only take nearest-neighbour coupling from $A$ to $B$ sites within the same layer or between 
the two different layers into account~\cite{couplings}. The four eigenenergies of the Hamiltonian (\ref{Ham}) are given by
\begin{eqnarray}
\nonumber \eps_{1,2} & = & \frac{U_1 + U_2}{2} + 
\frac{1}{2} \sqrt{ f_{\pm}(k) } \\
\eps_{3,4} & = & \frac{U_1 + U_2}{2} - \frac{1}{2} \sqrt{ f_{\pm}(k)}  
\end{eqnarray}
where $f_{\pm} (k) \equiv 4 k^2 + 2 t_{\perp}^2 + (U_1-U_2)^2$ $\pm$ $2 \sqrt{t_{\perp}^4 + 4 k^2 \left[ t_{\perp}^2 + (U_1-U_2)^2 \right] }$,
and $k$ $=$ $(k_x^2 + k_y^2)^{1/2}$ denotes the total momentum. From now onwards, we absorb a factor $(\hbar v)^{-1}$ in $\eps$, 
$U_1$, $U_2$ and $t_{\perp}$, which are all given in units of momentum.

Using scattering matrix theory we calculate in the next two sections the total pumped current~\cite{rectification} $I_p$ 
for $U_1=U_2\equiv U$ at the Dirac point $\eps=0$ and derive analytic expressions for the limit of heavily doped 
contacts $U \rightarrow -\infty$ (these were also considered in Ref.~\cite{snym07}).

\section{Calculations}
\label{calculations}

The scattering matrix $S$ and subsequently its derivatives with respect to $U_1$ and $U_2$ can be found by matching eigenstates 
of the Hamiltonian (\ref{Ham}) at the interfaces $x=0$ and $x=L$. For given $\eps$ and transverse momentum $k_y$ the eigenstates
of (\ref{Ham}) are characterized by two longitudinal momenta $k_{x\pm}$
\be
k_{x\pm} = \sqrt{ \frac{V_1^2 + V_2^2}{2} \pm \frac{1}{2} \sqrt{(V_1^2-V_2^2)^2 + 4 t_{\perp}^2 V_1 V_2} - k_y^2}
\label{veeka}
\ee
with $V_j \equiv \epsilon - U_j$ ($j=1,2$). Associated with each real wavevector $k_{x+}$ are two propagating modes 
$\phi_{\eps,+}^{R}(x,y)$ (right-going) and $\phi_{\eps,+}^{L}(x,y)$ (left-going). Similarly, another two propagating modes 
$\phi_{\eps,-}^{R}(x,y)$ and $\phi_{\eps,-}^{L}(x,y)$ correspond to each real wavevector $k_{x-}$. Defining $k_{\pm} \equiv
k_{x\pm} + i k_y$, the left- and right-going 
eigenstates are given by
\begin{subequations}
\bea
\phi_{\eps,\pm}^{R}(x,y) & = & N_{\pm} \left( \ba{c}
X_{2\pm} V_1 \\
X_{2\pm} k_{\pm}^{*} \\
X_{1\pm} V_2 \\
X_{1\pm} k_{\pm} 
\ea \right) e^{ik_{x\pm} x + i k_y y} 
\\
\phi_{\eps,\pm}^{L}(x,y) & = & N_{\pm} \left( \ba{c}
X_{2\pm} V_1 \\
- X_{2\pm} k_{\pm} \\
X_{1\pm} V_2 \\
- X_{1\pm} k_{\pm}^{*}
\ea \right) e^{-ik_{x\pm} x + i k_y y}
\eea
\label{eq:eigenstates}
\end{subequations}
where
\bea
X_{j\pm} & \equiv & V_j^2 - k_{x\pm}^2 - k_y^2 + t_{\perp} V_j \hspace*{1.cm} j=1,2 
\label{eq:X}\\
N_{\pm} & = & [2Wk_{x\pm} (V_1 X_{2\pm}^2 + V_2 X_{1\pm}^2) ]^{-1/2}. \label{eq:N}
\eea
The eigenstates (\ref{eq:eigenstates}) are normalized by $N_{\pm}$ such that each state carries unit current
\be
\frac{I}{ev} = \int_0^W dy\, \phi^{\dagger} \left( \ba{cc}
\sig_x & 0 \\
0 & \sig_x
\ea \right) \phi \equiv 1.
\ee
For $U_1 = U_2 \rightarrow -\infty$ Eqns.~(\ref{veeka}) and (\ref{eq:eigenstates}) reduce to the results of Ref.~\cite{snym07}.

From now onwards we assume to be at the Dirac point $\eps =0$. In the undoped graphene region $0<x<L$ we then
find from Eq.~(\ref{veeka}) that $k_{x\pm} = \pm i k_y$, which corresponds to evanescent modes. The left-incident 
eigenstates of the Hamiltonian (\ref{Ham}) can then be written as
\be
\psi_{\pm}(x,y) = \left\{ \ba{l}
\phi_{\eps,\pm}^R(x,y) + r_{+\pm}^{LL}\, \phi_{\eps,+}^L (x,y) +  r_{-\pm}^{LL}\, \phi_{\eps,-}^L (x,y) \\ 
\hspace*{4.cm} \mbox{\rm for}\ x<0 \\
\left[ (c_{1 \pm}\, \chi_1 + c_{2 \pm}\, \chi_2)\, e^{k_y x} + \right. \\ 
\left. \hspace*{0.5cm} (c_{3 \pm}\, \chi_3 + c_{4 \pm}\, \chi_4)\, e^{-k_y x} \right] e^{ik_y y} \\ 
\hspace*{4.cm} \mbox{\rm for}\ 0 < x < L \\
t_{+\pm}^{RL}\, \phi_{\eps,+}^R (x-L,y) + t_{-\pm}^{RL}\, \phi_{\eps,-}^R (x-L,y) \\
\hspace*{4.cm} \mbox{\rm for}\ x>L
\ea \right.
\label{eq:matching}
\ee
Here $r_{-+}^{\al \beta}$ and $t_{-+}^{\al \beta}$ with $\al,\beta \in \{L,R\}$ denote the reflection- and transmission
coefficients from a +-mode incident from lead $\beta$ to a $-$-mode in lead $\al$. In the middle region the eigenvectors 
$\chi_1$-$\chi_4$ have been constructed such that they are linearly independent at the Dirac point $\eps =0$. This yields
\bea
\chi_1 & = & \left( \ba{c}
0 \\ 1 \\ 0 \\ 0 \ea \right) \ \ , \ \
\chi_2 = \left( \ba{c}
0 \\ -it_{\perp}x \\ 1 \\ 0 \ea \right), \nn \\
\chi_3 & = & \left( \ba{c}
1 \\ 0 \\ 0 \\ -it_{\perp}x \ea \right)\ \ ,\ \ 
\chi_4 = \left( \ba{c}
0 \\ 0 \\ 0 \\ 1 \ea \right).
\eea
The reflection and transmission coefficients are calculated by matching the eigenstates (\ref{eq:matching}) at the two
interfaces $x=0$ and $x=L$, see Appendix~\ref{appendix}.
For a short and wide geometry with $L\ll W$ the boundary conditions in the $y$-direction become irrelevant. Taking
infinite mass boundary conditions, such that $k_y$ is quantized as $k_y=(n+1/2)\pi/W$, $n=0, 1, 2, \dots$
the pumped current (\ref{eq:current}) becomes a sum of eight terms:
\be
I_p  =  \frac{\omega e W \sin \phi\,  \delta U_1 \delta U_2}{2 \pi^2} \, 
\int_0^{\infty} dk_y \sum_{\sig, \sig^{\prime} = \pm}\, \Pi_{\sig \sig^{\prime}}(k_y)
\label{eq:current2}
\ee
with
\be
\Pi_{\sig \sig^{\prime}}(k_y) \equiv 
\mbox{\rm Im}\ \left(
\frac{\partial r_{\sig \sig^{\prime}}^{*LL}}{\partial V_1} \frac{\partial r_{\sig \sig^{\prime}}^{LL}}{\partial V_2} 
+ \frac{\partial t_{\sig \sig^{\prime}}^{*LR}}{\partial V_1} \frac{\partial t_{\sig \sig^{\prime}}^{LR}}{\partial V_2} 
\right).
\label{eq:Emis}
\ee

\section{Results}
\label{results}

In Appendix~\ref{appendix} we describe details of the calculation of the derivatives of 
$r_{\sig \sig^{\prime}}^{LL}$ and $t_{\sig \sig^{\prime}}^{LR}$ 
with respect
to $U_1$ and $U_2$ and consider the limit $U_1 \rightarrow U_2 \equiv U$ with
$U \rightarrow - \infty$. This is equivalent to $V_1 \rightarrow V_2 \equiv V$ with
$V \rightarrow + \infty$. Substitution of these derivatives into Eq.~(\ref{eq:current2}) and integrating over $k_y$ 
then yields the pumped current:
\begin{subequations}
\bea
I_p & = & \frac{\omega e W \sin \phi\,  \delta U_1 \delta U_2}{2 \pi^2} \, \frac{4 \lam}{U^2}\, \int_0^{V \rightarrow \infty} dk_y  \nn \\ 
& & \hspace{-0.2in} \frac{\left( \lam^2 \hspace{-0.03in} + \hspace{-0.03in} 6 k_y L \sinh (2 k_y L) \hspace{-0.03in} - \hspace{-0.03in} 8 \cosh^2(k_y L) \right)
\sinh(2 k_y L)}{\left(\lam^2 + 4 \cosh^2 (k_y L) \right)^3} \nn \\
\vspace{-0.2in} \label{eq:finalcurrent1}
 \\
& = & \frac{2\omega e}{\pi^2}\,  \frac{W}{L}\, \lam\, \sin \phi\, \frac{\delta U_1 \delta U_2}{U^2}\,
\int_0^{VL \rightarrow \infty}\, dx \nn \\
& & \frac{ \left( \lam^2 + 6 x \sinh(2x) - 8 \cosh^2(x) \right)
\sinh(2x)}{\left( \lam^2 + 4 \cosh^2(x) \right)^3} 
\label{eq:finalcurrent2} \\
& = & \frac{\omega e}{2\pi^2}\, \frac{W}{L}\, \sin \phi\, \frac{\delta U_1 \delta U_2}{U^2}\, 
\frac{1}{4\lam^2 (\lam^2 +4)^2}\,  \nn \\
& & \bigg( \lam^5 - 20 \lam^3 + 3 \lam (\lam^2+2) (\lam^2+4) \ln(\lam^2 + 4)    \nn \\
& & + 12 \sqrt{\lam^2 + 4} \Big[ Li_2 \big( 2 + \frac{\lam^2}{2} + \frac{\lam}{2} \sqrt{\lam^2 + 4} \big) \nn \\
& & - Li_2 \big( 2 + \frac{\lam^2}{2} - \frac{\lam}{2} \sqrt{\lam^2 + 4} \Big) \Big] \bigg).
\label{eq:finalcurrent3}
\eea
\label{eq:finalcurrent}
\end{subequations}
\noi Here we have defined the dimensionless coupling length $\lam \equiv t_{\perp} L$ and
$Li_2$ is the dilogarithm function defined as $Li_2 = \int_1^x (\ln t)/(1-t)\, dt$. 
Eq.~(\ref{eq:finalcurrent}) is the main result of this paper. The upper integration limit in Eq.~(\ref{eq:finalcurrent1}) 
originates from the requirement that the wavefunctions 
in the left and right leads should correspond to traveling waves, and hence that $k_{x\pm}$ should be real
($k_{x-}$ gives the more stringent condition $k_y \leq \sqrt{V(V-t)} \approx V$ for $V\rightarrow \infty$). 
We see that both for $L=0$ and for $L\rightarrow \infty$ the pumped current~(\ref{eq:finalcurrent3})
reduces to zero. For $L=0$, i.e. in the absence of the middle region, there is no possibility for 
evanescent waves to interfere in this region: all incoming waves from one lead are fully transmitted
into the other lead without scattering to the other layer and hence do not contribute to the pumped
current (which is composed of waves that scatter at least once from one layer to the other). 
For $L\rightarrow \infty$, the evanescent modes in the weakly-doped middle layer do not reach
the other contact and the pumped current becomes zero.

\begin{figure}[h]
\vspace{0.02in}
\psfrag{si}{$\; \sum_{\sigma \sigma^{\prime}=\pm} \Pi_{\sigma \sigma^{\prime}} \left[\frac{4 \lambda}{U^2} \right]$}
\psfrag{al}{$k_y L$}
\centering
\includegraphics[width=0.9\columnwidth]{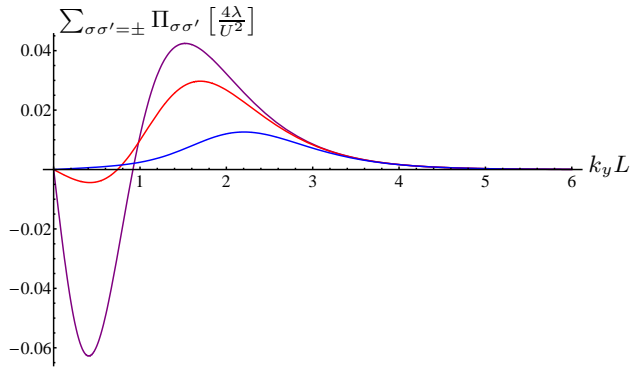}
\caption[]{The mode-dependent pumped current Eq.~(\ref{eq:emissivity}) as a function of $k_y L$. 
The curves correspond to $\lambda=0.01$ (purple), $\lambda=2$ (red), and $\lam=5$ (blue).
} 
\label{fig:current_vs_ky}
\end{figure} 
Fig.~\ref{fig:current_vs_ky} shows $\sum_{\sig, \sig^{\prime} = \pm}\, \Pi_{\sig \sig^{\prime}}(k_y)$
[Eq.~(\ref{eq:emissivity}) in units of $4\lam U^{-2}$ ], which is essentially the mode-dependent 
pumped current, as a function of $k_y$. We see that $\sum_{\sig, \sig^{\prime} = \pm}\, 
\Pi_{\sig \sig^{\prime}}(k_y=0)=0$,
i.e. waves with transverse momentum $k_y=0$ that are incident perpendicular to the interface  
do not contribute to the pumped current, which is a manifestation of the Klein paradox~\cite{cast09,kats06}. 
We also see that negative mode-contributions 
only occur for small values of $\lam$. From Eq.~(\ref{eq:emissivity}) it follows that 
the largest contribution to $\sum_{\sig, \sig^{\prime} = \pm}\, \Pi_{\sig \sig^{\prime}}$ comes
from the transverse modes $k_y L \sim 0.1\, \lam$ for $\lam
\sim 50$, similarly as for the transmission (conductance) in the same system~\cite{snym07}. 
\begin{figure}[h]
\vspace{0.02in}
\psfrag{ri}{$\; I_p \left[ \frac{\omega e}{2 \pi^2} \frac{W}{L} \frac{\delta U_1 \delta U_2}{U^2} \sin \phi \right]$}
\psfrag{am}{$\lambda$}
\centering
\includegraphics[width=0.9\columnwidth]{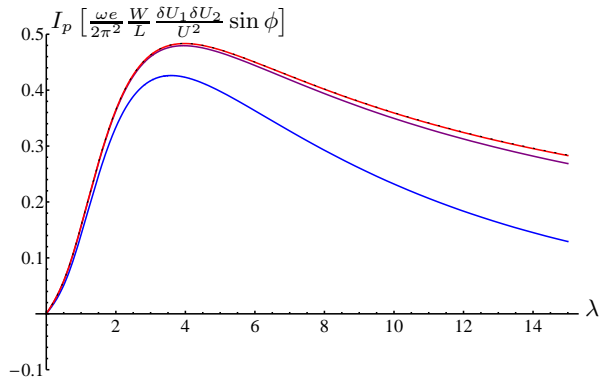}
\caption[]{The total pumped current $I_p$ as a function of $\lam$, for $W/L=100$. The blue, purple and red lines 
represent the exact sum over $k_y$ of the integrand in Eq.~(\ref{eq:finalcurrent1}) for
$n$=100, 150 and 200 modes. The black dotted line is the integrated result Eq.~(\ref{eq:finalcurrent3}).
} 
\label{fig:current_vs_lam}
\end{figure}   
Fig.~\ref{fig:current_vs_lam} shows a plot of the total pumped current $I_p$ as a function of $\lam$, for a fixed value of $W/L=100$. 
For $n \sim 200$ modes the sum converges to the continuum result. The number of modes needed for convergence becomes larger for $W/L$ larger, while for $W/L=20$ only $n \sim 20$ modes are needed.  
As expected, $I_p(t_{\perp}=0)=0$, since in the absence of interlayer coupling the bilayer reduces to two 
uncoupled monolayers for which the pumping parameters $V_1(t)$ and $V_2(t)$ become uncoupled and no pumping 
occurs. From Eq.~(\ref{eq:finalcurrent3}) 
we obtain that for small interlayer coupling strength $t_{\perp} \ll 1/L$ the pumped current $I_p$ scales as
\be 
\lim_{t_{\perp} \rightarrow 0} I_p = \frac{4 \ln 2 - 1}{64 \pi^2}\, \omega e\,
\frac{W}{L}\, \sin \phi\, \frac{\delta U_1 \delta U_2}{U^2}\, \lam,
\ee
hence $I_p$ depends linearly on $t_{\perp}$ for $t_{\perp} = 1/l_{\perp} \ll 1/L$. On the other
hand, in the limit of large interlayer coupling $t_{\perp} = 1/l_{\perp} \gg 1/L$ we find
\be
\lim_{t_{\perp} \rightarrow \infty} I_p = \frac{3}{4 \pi^2}\, \omega e\,
\frac{W}{L}\, \sin \phi\, \frac{\delta U_1 \delta U_2}{U^2}\, \frac{\ln \lambda}{\lam},
\ee
and hence $I_p \sim \frac{\ln t_{\perp}L}{t_{\perp}L}$ in this limit. In between these two limits
$I_p$ exhibits a maximum which is determined by $d I_p/d\lambda=0$ and yields 
\be
\lam_{\rm max} = 3.88\ \ \rightarrow \ \ I_p(\lam_{\rm max}) = 0.51\, \frac{\omega e}{2 \pi^2}\,
\frac{W}{L}\, \sin \phi\, \frac{\delta U_1 \delta U_2}{U^2}.
\ee
Thus $I_p$ is maximal if $t_{\perp}$ is of the same order as $1/L$. The pumped
current thus strongly depends on $t_{\perp}$ and reduces to zero for $t_{\perp} \rightarrow 0$.
This is in sharp contrast with the behavior of the conductance at the Dirac point in a graphene
bilayer, which equals the conductance across two monolayers and is independent of 
$t_{\perp}$~\cite{snym07}.

\section{Conclusion}
\label{conclusion} 

In summary, we have investigated the adiabatically pumped current through a wide graphene
bilayer consisting of a central weakly doped graphene sheet coupled to two heavily doped
contact regions. At the Dirac point, the pumped current is carried by evanescent waves
in the central region and exhibits a cross-over from linear (for $t_{\perp} L \ll 1$)
to logarithmic (for $t_{\perp} L \gg 1$)-dependence as a function of increasing interlayer
coupling strength $t_{\perp}$, with a maximum around $t_{\perp} L \sim 3.88$. This scaling
behavior with $t_{\perp}$ is markedly different from the behavior of the conductance $G$ in the
same system, which is independent of $t_{\perp}$ and equal to the conductance across two
monolayers in parallel. In practice, this different behavior of $I_p$ and $G$ as a function 
of $t_{\perp}$ and $L$ could be used to distinguish between the conductance and the pumped current.

We can estimate the magnitude of the pumped current using typical experimental 
parameters~\cite{cast09,cast10} $t_{\perp} = (0.4 eV)/(\hbar v_F) \approx
0.6\, 10^9$ m$^{-1}$, $U \approx 0.1$ V, $\delta U \approx 10$ mV, and $\omega \approx 1$ GHz.
In order to be able to measure a substantial pumped current, one thus needs sample sizes 
$L \sim 4/t_{\perp} \sim 6$ nm, which is smaller than currently available samples
of order $\mu$m~\cite{crac09}. However, with steady progress towards
smaller sample sizes, sample lengths of order 10-100 nm 
are expected to become available in the future.
We then obtain from Eq.~(\ref{eq:finalcurrent3}):
\be
I_{p,max} \sim 5 \cdot 10^{-14} \frac{W}{L} \sin(\phi)\, A
\ee
which, for $W/L \sim 100-1000$ is well within experimental reach. Observation of this current would be a striking
demonstration of quantum pumping produced by relativistic quantum mechanics.

\acknowledgements

This work has been supported by the Netherlands Organisation for
Scientific Research (NWO). 

\appendix
\section{Calculation of the scattering matrix elements and their derivatives}
\label{appendix}

In this appendix we calculate the reflection and transmission coefficients $r_{\sig \sig^{\prime}}^{\al \beta}$ and 
$t_{\sig \sig^{\prime}}^{\al \beta}$ ($\sig, \sig^{\prime} \in \{+,-\}$; $\al, \beta$ $\in$ {$L,R$}) from 
Eq.~(\ref{eq:matching}) and their derivatives with respect
to $U_1$ and $U_2$. These derivatives are then used to calculate the pumped current (\ref{eq:current2}).

Matching the eigenfunctions (\ref{eq:matching}) at the interfaces $x=0$ and $x=L$ results in the 8 equations:
\bea
&& r_{+\pm} N_{+} \left( \ba{c}
X_{2+} V_1 \\ - X_{2+} k_{+}  \\ X_{1+} V_2 \\ - X_{1+} k_{+}^{*} 
\ea \right)  +   r_{-\pm} N_{-} \left( \ba{c}
X_{2-} V_1 \\ - X_{2-} k_{-}  \\ X_{1-} V_2 \\ - X_{1-} k_{-}^{*} 
\ea \right) \nn \\
&& \hspace*{1.cm} = 
\left( \ba{c}
c_{3\pm} \\ c_{1\pm} \\ c_{2\pm} \\ c_{4\pm} 
\ea \right) 
- N_{\pm}\, \left( \ba{c}
X_{2\pm} V_1 \\ X_{2\pm} k_{\pm}^{*}  \\ X_{1\pm} V_2 \\ X_{1\pm} k_{\pm} 
\ea \right)  
\label{eq:match1}
\eea
\bea
&& t_{+\pm} N_{+} \left( \ba{c}
X_{2+} V_1 \\  X_{2+} k_{+}^{*}  \\ X_{1+} V_2 \\ X_{1+} k_{+} 
\ea \right) +  t_{-\pm} N_{-} \left( \ba{c}
X_{2-} V_1 \\ X_{2-} k_{-}^{*} \\ X_{1-} V_2 \\ X_{1-} k_{-} 
\ea \right) \nn \\
& & \hspace*{1.cm} =
\left( \ba{c}
c_{3\pm} z^{-1} \\ (c_{1\pm} -i t_{\perp} L c_{2\pm}) z\\ c_{2\pm} z \\ (c_{4\pm} -i t_{\perp} L c_{3\pm}) z^{-1}
\ea \right), 
\label{eq:match2}
\eea
with $z \equiv \exp(k_y L)$, $V_j = - U_j$ at the Dirac point ($j=1,2$) and $X_{j\pm}$ and $N_{\pm}$ given by 
Eqns.~(\ref{eq:X}) and (\ref{eq:N}).
Eliminating $c_1-c_4$ from Eqns.~(\ref{eq:match1}) and (\ref{eq:match2}) yields after some straightforward but lengthy algebra
for the four reflection coefficients:
\be
r_{\sig \sig^{\prime}}^{LL}  =  \frac{\al_{\sig \sig^{\prime}}^{LL} \cosh(2k_yL) + \beta _{\sig \sig^{\prime}}^{LL} \sinh(2k_yL) + \gamma _{\sig \sig^{\prime}}^{LL}}
{\delta \cosh(2k_yL) + \eps \sinh(2k_yL) + \eta} 
\label{eq:reflectioncoefficients}
\ee
with $\sig, \sig^{\prime} \in \{+,- \}$. Using $k_{\pm} \equiv k_{x\pm} + i k_y$, the dimensionless coupling length $\lam \equiv t_{\perp} L$,
$A \equiv X_{1+} X_{2-}$ and $B \equiv X_{1-} X_{2+}$, the coefficients in Eq.~(\ref{eq:reflectioncoefficients}) are
given by:
\bea
\al_{++}^{LL} & = & 2\, N_{+}\,N_{-}\, \left[ i k_y (k_{+} A^2   - k_{+}^{*} B^2)\,
+ \right. 
\nn \\
& & \left.  (k_{x+}^2 -  k_{x-}^2 + 2 k_y^2)\, A B  \right] \nn \\
\beta_{++}^{LL} & = & 2\, N_{+}\,N_{-} k_{x-}\, \left[ k_{+} A^2   - 
k_{+}^{*} B^2  - 2 i k_{y} A B  \right] \nn \\
\gamma_{++}^{LL} & = & N_{+}\,N_{-}\, \left[ \lam^2 (A-B)^2 V_1 V_2\, -\, \right. \nn \\
& & \hspace{-0.1in} \left. 2i \lam (A \hspace{-0.02in} - \hspace{-0.02in} B)  k_{x-} (X_{1+} X_{1-} V_2 \hspace{-0.02in} - \hspace{-0.02in} X_{2+} X_{2-} V_1) 
\right] - \al_{++}^{LL} \nn \\
\al_{+-}^{LL} & = & 2\, N_{-}^2\, k_{x-}  X_{1-} X_{2-} \left[ (k_{x+}-k_{x-} + 2 i k_y) A \right. \nn \\
& & \left. +\, (k_{x+} - k_{x-} - 2 i k_y) B 
\right] \label{eq:gamma++} \\
\beta_{+-}^{LL} & = & 2\, N_{-}^2 k_{x-}  X_{1-} X_{2-}  (k_{x+}+k_{x-}) (A-B)  \nn \\
\gamma_{+-}^{LL} & = & - 2 i \lam\, N_{-}^2\,(A-B)\,  k_{x-}  (X_{1-}^2 V_2 - X_{2-}^2 V_1) - \nn \\
& & \al_{+-}^{LL} \nn
\eea
and 
\begin{subequations}
\bea
\al_{-+}^{LL} & = & \al_{+-}^{LL} (\mbox{\rm \small subindex +} \leftrightarrow \mbox{\rm \small subindex -}) \nn \\
\beta_{-+}^{LL} & = &  \beta_{+-}^{LL} (\mbox{\rm \small subindex +} \leftrightarrow \mbox{\rm \small subindex -}) \nn \\
\gam_{-+}^{LL} & = &  \gam_{+-}^{LL} (\mbox{\rm \small subindex +} \leftrightarrow \mbox{\rm \small subindex -})\nn \\
\al_{--}^{LL} & = & \al_{++}^{LL} (\mbox{\rm \small subindex +} \leftrightarrow \mbox{\rm \small subindex -}) \nn \\
\beta_{--}^{LL} & = &  \beta_{++}^{LL} (\mbox{\rm \small subindex +} \leftrightarrow \mbox{\rm \small subindex -}) \nn \\
\gam_{--}^{LL} & = &  \gam_{++}^{LL} (\mbox{\rm \small subindex +} \leftrightarrow \mbox{\rm \small subindex -}), \nn
\eea
\label{eq:al-gamma2}
\end{subequations}
Also
\begin{subequations}
\bea
\del & = & 2\, N_{+}\,N_{-}\, \left[ -( k_{x+}k_{x-} - k_{y}^2)(A^2 + B^2)
\right. \nn \\
& & \left. +\, (k_{x+}^2 + k_{x-}^2 - 2 k_{y}^2) A B \right] \\
\eps & = & 2 i\, N_{+}\,N_{-}\, \left[ -(k_{x+}+ k_{x-})k_y\, (A^2  + B^2)
\right. \nn \\ 
& & \left. +\, 2\,(k_{x+} + k_{x-}) k_y\, A B \right] \\
\eta & = & N_{+}\,N_{-}\, \left[ -\lam^2 (A-B)^2 V_1 V_2 \right. \nn \\ 
& & - 2 i \lam   (k_{x+}-k_{x-}) (A-B) (X_{1+} X_{1-} V_2 - 
X_{2+} X_{2-} V_1) \nn \\ 
& & \left. - 2 (k_{x+} k_{x-} \hspace{-0.02in} +  \hspace{-0.02in} k_{y}^2) (A \hspace{-0.02in} - \hspace{-0.02in} B)^2 
 \hspace{-0.02in} - \hspace{-0.02in} 2(k_{x+} \hspace{-0.02in} - \hspace{-0.02in} k_{x-})^2 A B \right]
\eea
\label{eq:del-eta}
\end{subequations}
The transmission coefficients in Eq.~(\ref{eq:match2}) are given by
\begin{subequations}
\bea
t_{++}^{RL} & = & 
\frac{ \mu_{++}^{RL} \cosh(k_y L) - 
\nu_{++}^{RL} \sinh(k_yL)}
{\delta \cosh(2k_yL) + \eps \sinh(2k_yL) + \eta} \\
t_{+-}^{RL} & = & \frac{ \mu_{+-}^{RL} \cosh(k_y L) - 
\nu_{+-}^{RL} \sinh(k_yL)}
{\delta \cosh(2k_yL) + \eps \sinh(2k_yL) + \eta} \\
t_{-+}^{RL} & = & t_{+-}^{RL} (\mbox{\rm \small subindex +} \leftrightarrow \mbox{\rm \small subindex -}) \\
t_{--}^{RL} & = & t_{++}^{RL} (\mbox{\rm \small subindex +} \leftrightarrow \mbox{\rm \small subindex -}).
\eea
\label{eq:transmissioncoefficients}
\end{subequations}
with 
\bea
\mu_{++}^{RL} & = & - 2\, N_{+}\,N_{-}\,(A  -  B)\, k_{x+}\, \left[ 2 k_{x-} (A-B)
\right. \nn \\ & & 
\left. +\, i \lam (X_{1+} X_{1-} V_2 - X_{2+} X_{2-} V_1) \right] \nn \\
\nu_{++}^{RL} & = & 4 i\, N_{+}\,N_{-}\,(A-B)^2\, k_{x+} k_y  \nn \\
\mu_{+-}^{RL} & = & - 2 i\, \lam\, N_{-}^2\,(A-B)\, k_{x-} (X_{1-}^2 V_2 - X_{2-}^2 V_1) \nn \\
\nu_{+-}^{RL} & = & 2\, N_{-}^2\,(A-B)\, k_{x-} \left[ i \lam (X_{1-}^2 V_2 + X_{2-}^2 V_1) \right. \nn \\
& & \left. - 2\, X_{1-} X_{2-} (k_{x+} - k_{x-}) \right]
\eea
In two limiting cases the reflection- and transmission coefficients (\ref{eq:reflectioncoefficients}) 
and (\ref{eq:transmissioncoefficients}) reduce to simple forms
\vspace*{0.5cm}: \\
1) For $L\ria 0$, the $+$- and $-$-modes decouple and we obtain 
from Eq.~(\ref{eq:transmissioncoefficients}) 
\be
t_{++}^{RL} \ria 1\ \ \mbox{\rm and}\ \ t_{+-}^{RL} \ria 0.
\ee 
\\
2) For $t_{\perp} \ria 0$, i.e. in the absence of interlayer coupling, 
the reflection- and transmission coefficients  
(\ref{eq:reflectioncoefficients}) and (\ref{eq:transmissioncoefficients}) 
reduce to the monolayer expressions~\cite{kats06,twor06}
\begin{subequations}
\bea
r_{++}^{LL} & \stackrel{t_{\perp} \ria 0}{\ria} & \frac{(k_{x+} - ik_y) \sinh(k_y L)}{k_{x+} \cosh(k_yL) + i k_y \sinh(k_y L)}
\nn \\
& \stackrel{|V_1|, |V_2| \ria \infty}{\ria} & \tanh(k_y L) \\
t_{++}^{RL} & \stackrel{t_{\perp} \ria 0}{\ria} & \frac{k_{x+}}{k_{x+} \cosh(k_yL) + i k_y \sinh(k_y L)}
\nn \\
& \stackrel{|V_1|, |V_2| \ria \infty}{\ria} & \frac{1}{\cosh(k_y L)} \\
t_{+-}^{RL} & \stackrel{t_{\perp} \ria 0}{\ria} & 0.
\eea
\label{eq:limits}
\end{subequations}
The reflection- and transmission coefficients for right-incident Dirac fermions $r_{\sig \sig^{\prime}}^{RR}$
and $t_{\sig \sig^{\prime}}^{LR}$ ($\sig, \sig^{\prime} \in \{ +,- \}$) are given by Eqns.~(\ref{eq:reflectioncoefficients})
and (\ref{eq:transmissioncoefficients}) by interchanging the layer subindices 1 and 2. So $r_{++}^{RR} =
r_{++}^{LL}$ (\small subindex 1 $\leftrightarrow$ subindex 2) etc. \normalsize

In order to calculate the pumped current (\ref{eq:current2}), we need the
derivatives of $r_{\sig \sig^{\prime}}^{LL}$ and $t_{\sig \sig^{\prime}}^{LR}$ with respect to $U_1$ and $U_2$.
These can be calculated by splitting the coefficients $\al_{++}^{LL}$ etc. in Eq.~(\ref{eq:reflectioncoefficients}) into real
and imaginary parts and differentiating each of these with respect to $V_1$ and $V_2$. The latter does
not make any difference for the current, since only products of derivatives with respect to $U_1$ and $U_2$ enter 
Eq.~(\ref{eq:current2}) which are the same as those of the corresponding derivatives with respect to $V_1$ and $V_2$ (since $V_j = - U_j$ for $\eps =0$, $j=1,2$).
The resulting expressions for the derivatives are rather lengthy and therefore not given here. For equal bias voltages $V_1 = V_2 \equiv V$
(as required for a true pumping process) and in the limit of $V\rightarrow \infty$, however, an analytic expression for the pumped
current $I_p$ [Eq.~(\ref{eq:current}) can be derived. This expression is obtained by first taking the limit $V_1 \rightarrow V_2$, 
for which $X_{1-} \rightarrow X_{2-} = 2 t_{\perp} V$ and $X_{1+} \rightarrow - X_{2+} = (V_1 - V_2) (V + \frac{t}{2}) \rightarrow 0$
(see Eq.~(\ref{eq:X})). We then expand all terms in the Eq.~(\ref{eq:Emis}) in orders of $V_1 - V_2$ (or, equivalently, $X_{1+}$) and find that while the numerators contain terms of order $(V_1 - V_2)^7$ and higher, all terms in the denominator are proportional to $(V_1 - V_2)^8$.
In the limit of $V \rightarrow \infty$, assuming $V \gg t_{\perp}, k_y$ and $V t_{\perp} \gg k_y^2$, the terms of order $(V_1 - V_2)^7$ in the numerator cancel and we are left with a finite
pumped current which to highest order in $V$ amounts to:
\bea
\sum_{\sig, \sig^{\prime} = \pm} & & \! \! \Pi_{\sig \sig^{\prime}}(k_y)  =  \frac{4 \lam}{U^2} \times \nn \\
& & \hspace{-0.2in} \frac{\left( \lam^2 \hspace{-0.02in} + \hspace{-0.02in} 6 k_y L \sinh (2 k_y L) \hspace{-0.02in} - \hspace{-0.02in} 8 \cosh^2(k_y L) \right)
\sinh(2 k_y L)}{\left(\lam^2 + 4 \cosh^2 (k_y L) \right)^3} \nn \\
\label{eq:emissivity}
\eea
Integration of (\ref{eq:emissivity}) over the transverse modes $k_y$ then leads to the pumped current Eq.~(\ref{eq:finalcurrent})
in the main text.


\begin{thebibliography}{99}
\bibitem{couplings}
Hamiltonians in which also different couplings were taken into account can 
be found in Ref.~\cite{cast09}.
\bibitem{rectification}
Our pumped current can also be considered as a rectified net dc current between the 
left and right leads that is induced by an ac voltage bias
between the top and bottom leads in the geometry of Fig.~\ref{fig:geometry}.
\end{thebibliography}
\end{document}